# Scattering Processes on the Interface Roughness in Quantum wells in a Tilted Quantizing Magnetic Field

M.P. Telenkov, Yu.A. Mityagin

P.N. Lebedev Physical Institute of the Russian Academy of Sciences, 119991, Moscow, Russia.

Scattering processes by the interface roughness in a quantum well in a quantizing magnetic field are considered. An expression for the scattering rate is derived for a magnetic field tilted relative to the quantum well layers. By analyzing this expression, trends in the behavior of the scattering rate are established with variation in the magnetic field strength and orientation, the potential profile of the quantum well, and the interface roughness parameters.

# 1. Introduction

A magnetic field applied perpendicular to the quantum well layers considerably modifies the nature of the quantum well's energy spectrum. Continuous two-dimensional subbands of quantum confinement become discrete series of Landau levels, each degenerate with a macroscopic multiplicity [1]. This change in the structure of the energy spectrum has a significant impact on the scattering and relaxation processes of electrons in quantum wells [2-25].

During the fabrication of quantum wells, roughness at the interfaces inevitably arises. For this reason, scattering processes from the interface roughness have been studied in considerable detail [26-45], including in a quantizing magnetic field [1,6,9,12,14,16,18,21].

However, scattering processes have been studied mostly in a quantizing magnetic field directed perpendicularly to the quantum well layers. At the same time, there are data [46–50] that by tilting the magnetic field relative to the plane of the quantum well structure, one can significantly influence various transitions between Landau levels.

In this article, expressions are derived for the scattering rate by interface roughness in a quantizing magnetic field tilted relative to the quantum well layers. The behavior of scattering processes by interface roughness is determined as the magnetic field strength and orientation vary, as well as the quantum well potential profile and interface roughness parameters.

# 2. Electron spectrum in a quantum well in a tilted magnetic field

In the magnetic field $\mathbf{B} = B_\perp \mathbf{e}_z + B_\parallel \mathbf{e}_y$ tilted at an angle $\theta$ to the z-axis of quantum well growth axis the Hamilton operator of the parabolic approximation of the envelope function formalism [51]

$$\hat{H} = \left(\hat{\mathbf{p}} + \frac{e}{c}\mathbf{A}\right)\frac{1}{2m(z)}\left(\hat{\mathbf{p}} + \frac{e}{c}\mathbf{A}\right) + U(z) \tag{1}$$

takes in the Landau gauge $\mathbf{A} = (B_\parallel z - B_\perp y)\mathbf{e}_x$, the form

$$\hat{H} = \hat{\mathbf{p}}\frac{1}{2m(z)}\hat{\mathbf{p}} + U(z) + \frac{m_w}{m(z)}\left[(\omega_\parallel z - \omega_\perp y)\hat{p}_x + \frac{m_w}{2}(\omega_\parallel z - \omega_\perp y)^2\right]. \tag{2}$$

Here $U(z)$ - the potential profile of the quantum well, $m(z)$ - the effective electron mass ($m_w$ in the well and $m_b$ in the barrier), $\omega_\perp = eB_\perp / m_w c$ and $\omega_\parallel = eB_\parallel / m_w c$ - the cyclotron frequencies for the magnetic field components along the growth axis of the structure ($B_\perp$) and in the plane of its layers ($B_\parallel$) respectively.

Since $\hat{H}\hat{p}_x - \hat{p}_x \hat{H} = 0$ then we can construct a basis of stationary states with a particular momentum projection value $p_x = \hbar k_x$ on the x-axis. The wave functions of such a basis have the form

$$\Psi(x,y,z) = \frac{\exp(ik_x x)}{\sqrt{L}} \psi(y - k_x \ell_\perp^2, z), \tag{3}$$

where $\ell_\perp = \sqrt{\hbar / m\omega_\perp} = \sqrt{\hbar c / eB_\perp}$ is the magnetic length for the transverse component of the magnetic field. The electron energy levels and wave functions of stationary states are defined by the two-dimensional Hamiltonian [52]

$$\hat{H}_{2D} = \hat{H}_\perp + \hat{H}_\parallel, \tag{4}$$

where

$$\hat{H}_\perp = -\frac{\partial}{\partial z}\frac{\hbar^2}{2m(z)}\frac{\partial}{\partial z} + U(z) + \frac{m_w}{m(z)}\left[\frac{\hat{p}_y^2}{2m_w} + \frac{m_w \omega_\perp^2}{2} y^2\right] \tag{5}$$

is the electron Hamiltonian in the magnetic field directed along the growth axis of the structure. The term

$$\hat{H}_\parallel = +\frac{m_w}{m(z)}\frac{m_w \omega_\parallel^2 z^2}{2} - \frac{m_w}{m(z)} m_w \omega_\parallel \omega_\perp z y \tag{6}$$

is due to the magnetic field component being parallel to the layers of the quantum well.

The variables in the Schrödinger equation with the Hamiltonian (7) are separable. The energy levels have the form

$$E_{(v,n)} = \varepsilon_v + \hbar\omega_c\left(n + \frac{1}{2}\right), \tag{7}$$

and the wave functions are given by the expression [53]

$$\Psi(x,y,z) = \frac{\exp(ik_x x)}{\sqrt{L}} \varphi_v(z) \Phi_n(y), \tag{8}$$

where L is the transverse dimension of the heterostructure, $\varphi_\nu(z)$ is the wave function of the subband level $\varepsilon_\nu$ (eigenvalue wave function of the Hamiltonian $\hat{H}_z = -\dfrac{\partial}{\partial z}\dfrac{\hbar^2}{2m(z)}\dfrac{\partial}{\partial z} + U(z)$), $\Phi_n(y)$ - the wave function of the n-th (n=0,1,2,...) energy level of a linear harmonic oscillator with cyclotron frequency $\omega_\perp$.

We neglect the effect of decreasing the barrier height with increasing Landau level number n [53] due to its smallness for the deep subbands under consideration.

The Hamiltonian matrix (4) in the basis of wave functions (5) is diagonal in $k_x$, and the matrix element at $k_{x1} = k_{x2}$

$$\left\langle \dfrac{\exp(ik_x x)}{\sqrt{L}} w_1(y - k_x \ell_\perp^2, z) \middle| \hat{H} \middle| \dfrac{\exp(ik_x x)}{\sqrt{L}} w_2(y - k_x \ell_\perp^2, z) \right\rangle = \\ = \left\langle w_1(y,z) \middle| \hat{H}_{2D} \middle| w_2(y,z) \right\rangle \qquad (9)$$

does not depend on $k_x$. Therefore, in a tilted quantizing magnetic field, the energy levels are degenerate in $k_x$ and the degeneracy factor is determined only by the component of the magnetic field $B_\perp$

$$\alpha = \dfrac{e}{\pi\hbar c} \cdot B_\perp = 4.9 \cdot 10^{10} \cdot B_\perp \, \dfrac{cm^{-2}}{T}. \qquad (10)$$

The matrix element between the Landau levels $(\nu_1, n_1)$ and $(\nu_2, n_2)$ is given by the expression [54]

$$\left\langle \nu_1, n_1 \middle| \hat{H}_{2D} \middle| \nu_2, n_2 \right\rangle = \left[ \varepsilon_{\nu_1} + \hbar\omega_\perp(n + 1/2) \right] \delta_{\nu_1,\nu_2} \delta_{n_1,n_2} + \\ + \dfrac{m_w \omega_\parallel^2}{2} \langle z^2 \rangle_{\nu_1,\nu_2} \delta_{n_1,n_2} - m_w \hbar\omega_\parallel \sqrt{\hbar\omega_\perp} \sqrt{\dfrac{m_w}{2\hbar^2}} \langle z \rangle_{\nu_1,\nu_2} \times \\ \times \left[ \sqrt{n_2 + 1} \cdot \delta_{n_1,n_2+1} + \sqrt{n_2} \cdot \delta_{n_1,n_2-1} \right] \qquad (11)$$

In a quantum well in a tilted magnetic field, there are two energy scales: the cyclotron energy and the quantum confinement energy, determined by the potential profile of the heterostructure (the distance between quantum confinement subbands). We are interesting in the case where the cyclotron energy is several times smaller than the quantum confinement energy. In this case, the coupling between subbands (elements with $v_1 \neq v_2$) in matrix (11) can be neglected and matrix can be diagonalized analytically [39,46,55,56]. As a result, the following expressions are obtained for the Landau levels and the wave functions of stationary states

$$E_{(v,n)} = \varepsilon_v + \Delta_v(B_\parallel) + \hbar\omega_\perp\left(n + \frac{1}{2}\right) \tag{12}$$

and

$$\psi_{(v,n),k_x}(x,y,z) = \frac{\exp(ik_x x)}{\sqrt{L}} \varphi_v(z) \Phi_n\left(y - k_x \ell_\perp^2 - \langle z \rangle_v tg\theta\right). \tag{13}$$

Here

$$\Delta_v(B_\parallel) = \frac{m_w \omega_\parallel^2}{2}(\delta z)_v^2 = \frac{e^2}{2m_w c^2}(\delta z)_v^2 \cdot B_\parallel^2, \tag{14}$$

Is the subband shift caused by the magnetic field component $B_\parallel$,

$$\langle z \rangle_v = \int dz\, \varphi^*(z) z \varphi(z) \tag{15}$$

is the average value of the z-coordinate of the electron, $(\delta z)_v$ is its standard deviation, $tg\theta = B_\parallel / B_\perp$.

We neglect the spin splitting of the Landau levels due to its small values in the considered structures made of III-V semiconductors of GaAs type. In the considered range of magnetic fields (1-10 T), the magnitude of the Zeeman splitting is significantly smaller than the width of the Landau level [5,57,58].

### 3. Interface roughness scattering rate

In this paper, we use the standard model of heterointerface roughness [27]. The heterointerface surface is described by its deviation $\eta(x,y)$ from the midplane $z = z_0$ (an ideal heterointerface). The function $\eta(x,y)$ is considered random with a mean value

$$\langle \eta \rangle = 0, \tag{16}$$

and autocorrelation function

$$\langle \eta(\mathbf{r}_\perp)\eta(\mathbf{r}'_\perp)\rangle = \eta_0^2 \exp\left(-\frac{|\mathbf{r}_\perp - \mathbf{r}'_\perp|^2}{\lambda^2}\right), \tag{17}$$

where $\langle\ldots\rangle$ means averaging over all possible configurations of the heterointerface, $\eta_0$ -

The contribution to the electron Hamiltonian due to the roughness of the heterointerface has the form [34]

$$\hat{H}^{rough} = U_0 \delta(z - z_0) \eta(x, y), \tag{18}$$

where $U_0$ - is the barrier height.

According to the Fermi rule, the scattering rate from Landau level $i = (v_i, n_i)$ to Landau level $f = (v_f, n_f)$ is

$$\frac{1}{\tau_{i\to f}} = A_{i\to f} \cdot \delta(E_f - E_i), \tag{19}$$

where

$$A_{i\to f} \frac{2}{\alpha L^2} \sum_{k_i, k_f} \langle |H_{f,i}^{rough}(k_f, k_i)|^2\rangle = \frac{1}{2\pi^2 \alpha} \int dk_i dk_f \langle |H_{f,i}^{rough}(k_f, k_i)|^2\rangle, \tag{20}$$

is a scattering amplitude,

$$H_{f,i}^{rough}(k_f, k_i) = \int d\mathbf{r}\, \psi_{f,k}^*(\mathbf{r}) \hat{H}_{rough} \psi_{i,k}(\mathbf{r}). \tag{21}$$

The Dirac delta function expresses the energy conservation law during scattering (the resonance condition of $i \to f$ transition):

$$E_f = E_i. \tag{22}$$

Using the wave functions (13), we obtain the following expression for the scattering amplitude in a tilted magnetic field

$$A_{i\to f}^{rough}(B_\parallel) = A_{i\to f}^{rough}(B_\parallel = 0) \cdot G_{n_i, n_f}\left(\frac{\xi_{v_f, v_i}^2}{2(2\gamma + 1)}\right), \tag{23}$$

where

$$A_{i \to f}^{rough}(B_{\parallel} = 0) = K_{v_i,v_f} \cdot Q_{n_i,n_f}(\gamma) \tag{24}$$

is the transition amplitude in the case where the magnetic field is directed perpendicular to the layers of the quantum well ($B_{\parallel} = 0$),

$$K_{v_i,v_f} = \frac{2\pi U_0^2 \eta_0^2}{\hbar} |\varphi_{v_i}(z_0)|^2 |\varphi_{v_f}(z_0)|^2, \tag{25}$$

$$Q_{n_i,n_f}(\gamma) = \frac{1}{(1+2\gamma)^{n_i+n_f+1}} \sum_{j=0}^{\min\{n_i,n_f\}} \binom{n_i}{j}\binom{n_f}{j} (2\gamma)^{n_i+n_f-2j}, \tag{26}$$

$$G_{n_i,n_f}(x) = \exp(-x) P_{n_i,n_f}(x), \tag{27}$$

$$P_{n_i,n_f}(x) = \frac{\int dp \exp\left(-\left(\frac{2\gamma+1}{4\gamma}\right)p^2\right) \cdot M_{i,f}\left(\frac{p}{2}+x\right)}{2^{n_i+n_f+1} n_i! n_f! \pi^{3/2} \gamma^{1/2} Q_{n_i,n_f}(\gamma)}, \tag{28}$$

$$M_{i,f}(x) = \int dy_1 \exp\left(-\frac{(2\gamma+1)}{(\gamma+1)} y_1^2\right) H_{n_i}(y_1+x) H_{n_f}(y_1-x) \times$$
$$\times I\left(\delta = (\gamma+1); \mu_1 = \frac{\gamma}{(\gamma+1)} y_1 + x; \mu_2 = \frac{\gamma}{(\gamma+1)} y_1 - x\right) \tag{29}$$

$$I(\delta;\mu_1;\mu_2) = \int dy \exp(-\delta y^2) H_{n_i}(y+\mu_1) H_{n_f}(y+\mu_2), \tag{30}$$

$$\gamma = \left(\frac{\ell_\perp}{\lambda}\right)^2, \tag{31}$$

$$\xi_{v_f,v_i} = \left[\langle z \rangle_{v_f} - \langle z \rangle_{v_i}\right] \frac{\ell_\perp}{\ell_\parallel^2} = \sqrt{\frac{e}{\hbar c B_\perp}} \left[\langle z \rangle_{v_f} - \langle z \rangle_{v_i}\right] B_\parallel^2. \tag{32}$$

## 4. Transitions due to scattering on the interface roughness

This section analyzes the obtained expressions, and the basic trends in the behavior of the scattering rate from the magnetic field components $B_\perp$ and $B_\parallel$ are established. The analysis will be accompanied by illustrations using the example of GaAs/Al0.3Ga0.7As quantum wells. The band parameters ($m_w = 0.067 m_0$, $m_b = 0.0913 m_0$, barrier height = 240 meV) are taken from [60].

The effect of the magnetic field components $B_\perp$ and $B_\parallel$ on the electron spectrum differs significantly.

The component $B_\perp$ leads to quantization of the electron energy. A magnetic field perpendicular to the layers transforms each continuous two-dimensional quantum-well subband into a discrete, equidistant set of Landau levels (7). The distance between Landau levels (the Landau energy) is proportional to the magnetic field component $B_\perp$ perpendicular to the quantum well layers. Each Landau level is macroscopically degenerate. Moreover, the degeneracy factor of a Landau level is also determined only by the magnetic field component $B_\perp$ perpendicular to the layers.

The discrete nature of the electron spectrum within a subband means that resonance condition (22) for intrasubband transitions is not satisfied at finite values of $B_\perp$. Since the distance between Landau levels increases with $B_\perp$, the quantizing magnetic field suppresses intrasubband imterface roughness scattering processes.

A different situation occurs for intersubband scattering. In this case, the discrete nature of the spectrum leads to the resonance condition being satisfied for a discrete set of values. When the magnetic field is directed perpendicular to the layers, resonance of the intersubband transition occurs at a magnetic field value

$$B_{\perp,\Delta n}^{(0)} = \frac{m_w c}{e\hbar} \frac{\Delta \varepsilon_{if}}{\Delta n}, \tag{33}$$

where

$$\Delta \varepsilon_{if} = \varepsilon_{v_i} - \varepsilon_{v_f} \tag{34}$$

is the intersubband distance in zero magnetic field,

$$\Delta n = n_f - n_i. \tag{35}$$

Therefore, the dependence of the transition rate $1/\tau_{(v_i,n_i) \to (v_f,n_f)}$ on the magnetic field $B_\perp$ has a resonant character.

The broadening of the Landau levels leads to the replacement of the Dirac delta function in (19 by a form factor of finite width. The form factor type and its width depend on a number of factors - the relations between collisional and inhomogeneous broadening [5], between collisional broadening and temperature [61-63], between the radius of the scatterers and the magnetic length (which is determined by the electron density and the degeneracy factor of the Landau levels) [1,64]. However, in any of these situations, the scattering rate has a maximum at a magnetic field value for which the energy conservation law is satisfied. In this paper, we study the effects associated with the influence of the magnetic field on the resonance position (22) and the transition amplitude $A^{rough}$, for which the specific form factor does not play a fundamental role. Therefore, in what follows, we replace the Dirac delta function in (32) with a Gaussian

$$F\left(E_i - E_f\right) = \frac{1}{\sqrt{2\pi}\left(\sqrt{2}\Gamma\right)} \exp\left(-\frac{\left(E_i - E_f\right)^2}{2\left(\sqrt{2}\Gamma\right)^2}\right) \qquad (36)$$

with a typical width of $\Gamma = 1$ meV.

We will further consider transitions from the upper subband to the lower one. In this case $\Delta\varepsilon_{if} > 0$. As it follows from (33), resonances occur for transitions with $\Delta n > 0$, i.e., for transitions in which the Landau level number $n_f$ in the final state exceeds the Landau level number $n_i$ in the initial state. Thus, the discrete set (33) of resonant values of the magnetic field is limited from above by the value for $\Delta n = 1$

$$B^{(0)}_{\perp,\Delta n} < B^{(0)}_{\perp,\Delta n=1} = \frac{m_w c}{e\hbar}\Delta\varepsilon_{if}. \qquad (37)$$

The set of resonant values is not limited from below - $\Delta n$ can take on an arbitrarily large value, and, accordingly, the resonant field (33) can be arbitrarily small. As a result, the total scattering rate from the Landau level $\left(v_i, n_i\right)$ to the underlying subband $1/\tau^{(tot)}_{(v_i,n_i)} = \sum_{n_f=n_i+1}^{\infty} 1/\tau_{(v_i,n_i)\to(v_f,n_f)}$ has an oscillatory behavior, turning into a monotonic decrease at $B_\perp > B^{(0)}_{\perp,\Delta n=1}$ (Fig. 1).

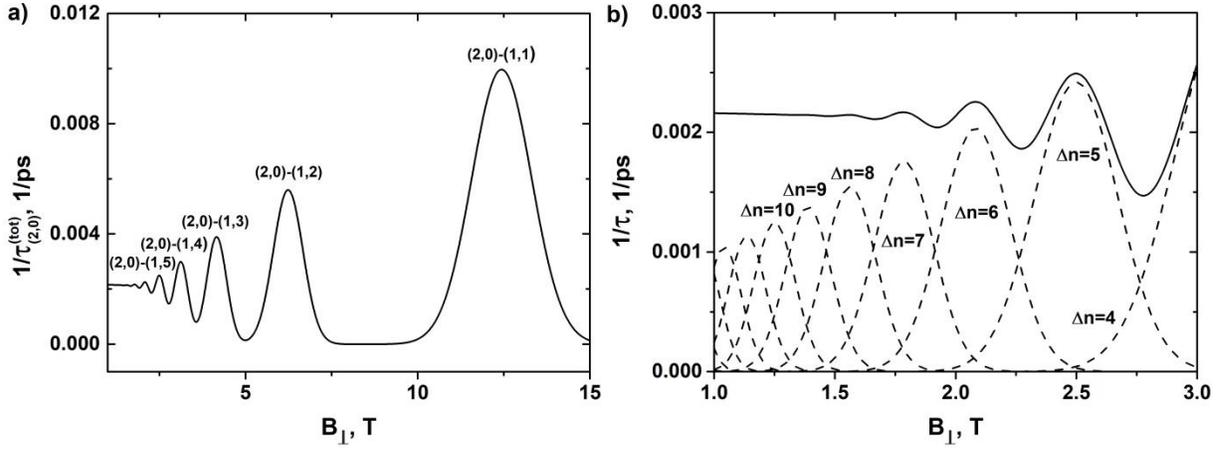

**Figure 1**. Magnetic field dependence of the total electron transition rate from the (2,0) level due to scattering by the interface roughness. The dashed lines indicate the transition rates from the (2,0) level to individual Landau levels (1, Δn) of the lower subband. The data are presented for a 25 nm wide $GaAs/Al_{0.3}Ga_{0.7}As$ quantum well.

Resonances of transitions with a large change in the Landau level number and, correspondingly, a larger difference in the wave functions of the initial and final states (in particular, the difference in the number of zeros of the wave functions of the initial and final states) arise in weaker magnetic fields. Consequently, the amplitude of the maxima of the total scattering rate decreases with decreasing magnetic field. The distance between the resonances of adjacent transitions and the difference in the magnitude of their maxima decrease with decreasing magnetic field. As a result, at relatively weak magnetic fields, the dependence smooths out (becomes weakly oscillatory) due to the summation of closely spaced peaks with similar amplitudes (Fig. 1b).

According to (12), the application of a magnetic field $\mathbf{B}_\parallel = B_\parallel \mathbf{e}_y$ parallel to the structure layers, in addition to the quantizing magnetic field, leads to a shift of each subband as a whole by an amount (14), proportional to $B_\parallel^2$ and the square of the $(\delta z)_\nu$ (root-mean-square deviation of the coordinate $z$ along the growth axis of the quantum well).

According to the oscillator theorem, the wave function $\varphi_\nu(z)$ has $\nu - 1$ zeros. Accordingly, with increasing $\nu$, the behavior of $\varphi_\nu(z)$ becomes more complex, leading to an increase in the standard deviation $(\delta z)_\nu$. Thus, the component of the magnetic field parallel to the layers leads to an increase in the distance between subbands. This increase in the intersubband distance leads to the dependence of the resonance condition (22) on the magnetic field component $B_\parallel$.

Substituting the explicit form of energy levels (12) into the resonance condition (22), we obtain

$$\Delta\varepsilon_{if} + \delta\varepsilon_{if}(B_\parallel) - \hbar\omega_\perp \Delta n = 0, \tag{38}$$

where

$$\delta\varepsilon_{if}(B_\parallel) = \frac{e^2}{2m_w c^2}\left[(\delta z)^2_{v_i} - (\delta z)^2_{v_f}\right] \cdot B_\parallel^2 \tag{39}$$

is an increment in the distance between subbands caused by the magnetic field component $B_\parallel$ parallel to the quantum well layers.

Condition (38) is satisfied in a magnetic field

$$B_\perp = B_\perp^{(0)}\left(1 + \frac{\delta\varepsilon_{if}(B_\parallel)}{\Delta\varepsilon_{if}}\right), \tag{40}$$

For transitions from the upper to the lower subband, both $\Delta\varepsilon_{if}$ and $\delta\varepsilon_{if}$ are positive. Therefore, applying a magnetic field $B_\parallel$ parallel to the layers leads to a shift of each resonance toward higher magnetic fields (Fig. 2) by

$$\delta B_\perp = B_\perp - B_\perp^{(0)} = \frac{m_w c}{e\hbar} \frac{\delta\varepsilon_{if}(B_\parallel)}{\Delta n}, \tag{41}$$

The relative magnitude of this shift for transitions between two given subbands

$$\frac{B_\perp - B_\perp^{(0)}}{B_\perp^{(0)}} = \frac{\delta\varepsilon_{if}(B_\parallel)}{\Delta\varepsilon_{if}}, \tag{42}$$

does not depend on the transition type and is equal to the relative change in the intersubband distance caused by the magnetic field component $B_\parallel$.

The indicated shifts of the resonance peaks towards larger values of the quantizing component of the magnetic field and the accompanying increase in amplitude are clearly visible in dependence of the total scattering rate $1/\tau_i^{(tot)}$ on $B_\perp$ from the Landau level in the region when these peaks are resolved (Fig. 2).

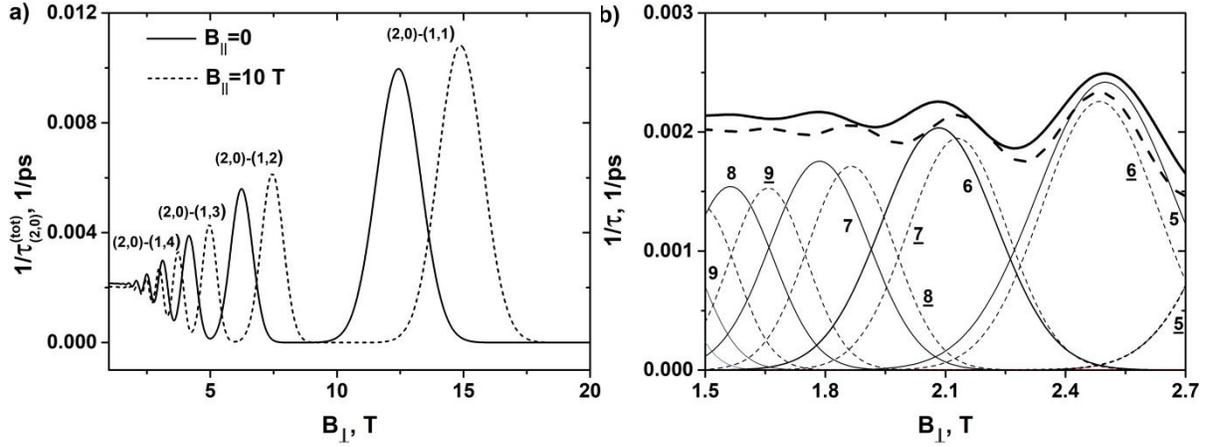

**Figure 2**. Total interface roughness scattering rate from the (2,0) level as a function of the quantizing magnetic field $B_\perp$ (solid curve) and a similar dependence upon adding a magnetic field $B_\parallel$ parallel to the quantum well layers (dashed curve). Curve number $n$ is the rate of an individual transition $(2,0) \to (1,n)$. The data are given for a GaAs/Al$_{0.3}$Ga$_{0.7}$As quantum well with a width of 25 nm and a magnetic field component parallel to the layers, $B_\parallel$=10 T.

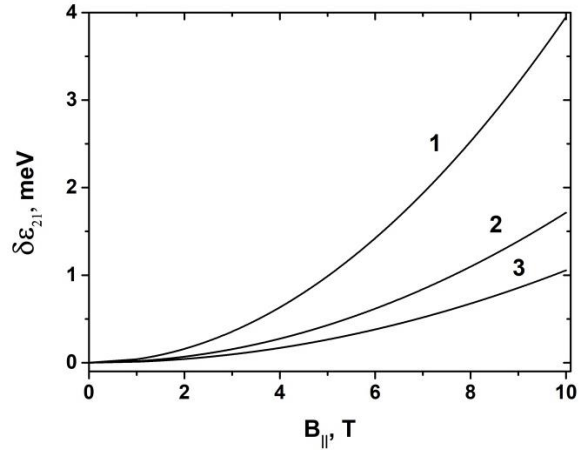

**Figure 3**. Dependence of the distance between two lower subbands of a quantum well on the magnetic field component parallel to its layers. Different curves correspond to different values of the width $a$ of the quantum well: 1 – $a$=25 nm; 2 – $a$=15 nm; 3 – $a$=10 nm. The data are given for a quantum well GaAs/Al$_{0.3}$Ga$_{0.7}$As.

In the region of small $B_\perp$, when the resonances of individual transitions overlap and the dependence becomes weakly oscillating, a decrease in the total rate $1/\tau_i^{(tot)}$ is observed with the addition of $B_\parallel$, despite an increase in the amplitude of the resonance peaks of individual transitions. The reason for this behavior is clearly visible in Figure 2b. The shift in resonances results in retaining transition resonances with smaller values $\Delta n$ and, consequently, smaller amplitudes $A_{i \to f}^{rough}$, in the region where the weakly oscillating dependence occurs. The summation of nearby resonances with smaller amplitudes results in a smaller value of $1/\tau_i^{(tot)}$.

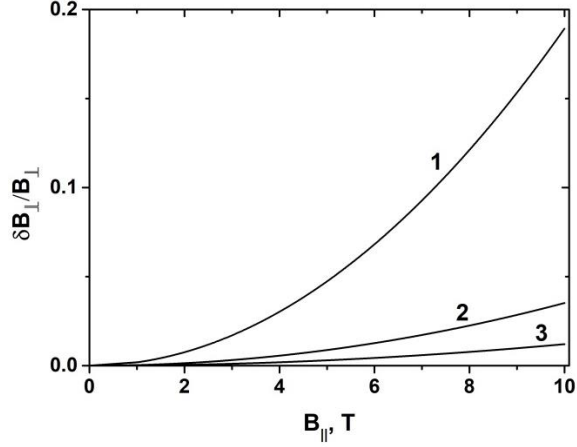

**Figure 4.** Dependence of the relative shift of the resonance position of the transition between two Landau levels of different subbands on the magnetic field component parallel to the quantum well layers. Different curves correspond to different values of the width $a$ of the quantum well: 1 – $a=25$ nm; 2 – $a=15$ nm; 3 – $a=10$ nm. The data are given for transitions from the Landau level of the second subband to the Landau level of the lower subband of the quantum well GaAs/Al$_{0.3}$Ga$_{0.7}$As.

The increment in the distance between the subbands $\delta\varepsilon_{if}(B_\parallel)$ increases with increasing quantum well width $a$ (Fig. 3). This is due to the increase of the localization length of the wave function $\varphi(z)$, which leads to an increase in the standard deviation $(\delta z)_v$. The form of this dependence can be determined by considering deep energy levels in a rectangular quantum well, the wave functions of which are localized predominantly in the quantum well ($\hbar/\sqrt{2m(U_0-\varepsilon)} \ll a$). In this case, one can estimate $\delta z$ and $\Delta\varepsilon_{if}$ using the energy levels and wave functions in an infinitely deep quantum well. In this case, we obtain

$$(\delta z)_v^2 = \frac{a^2}{12}\left(1 - \frac{6}{(\pi v)^2}\right). \qquad (43)$$

Correspondingly,

$$\delta\varepsilon_{if}\left(B_{\|}\right) = \frac{e^2}{4\pi^2 m_w c^2}\left(\frac{1}{v_f^2} - \frac{1}{v_i^2}\right)\cdot a^2 \cdot B_{\|}^2 \tag{44}$$

Thus, the distance between the subbands depends significantly on the quantum well width $a$ —it increases proportional to $a^2$. Since $\Delta\varepsilon_{if} \sim 1/a^2$, the relative magnitude of the shift (40) increases with the quantum well width approximately $\sim a^4$ (Fig. 4).

One can see from Figure 2, that the resonance shift caused by $B_{\|}$ is accompanied by a change in the maximum scattering rate $1/\tau_{i\to f}$. The nature of this change depends significantly on the type of transition and the symmetry of the quantum well potential profile. From the resulting expressions, it is immediately evident that the component of the magnetic field $B_{\|}$ parallel to the layers enters into expression (23) for the transition amplitude $A^{rough}$ only through the parameter $\xi_{v_f,v_i}$, defined by formula (32). When this parameter is zero, there is no dependence of $A^{rough}$ on $B_{\|}$.

At $B_{\|} \neq 0$, the parameter $\xi_{v_f,v_i}$ is zero if the average coordinates $\langle z \rangle$ in the initial and final states are equal, i.e., $\langle z \rangle_i = \langle z \rangle_f$.

This is certainly the case for all intrasubband transitions. Consequently, the amplitude of all intrasubband transitions is independent of the magnetic field component $B_{\|}$. The resonance condition for an intrasubband transition is also independent of the component parallel to the layers. Thus, we conclude that the magnetic field component parallel to the quantum well layers has no effect on intrasubband transitions.

The influence $B_{\|}$ on intersubband transitions is determined by the symmetry of the quantum well's potential profile $U(z)$.

In the case of a symmetric potential profile ($U(-z) = U(z)$), the wave functions of the quantum confinement levels are either even or odd

$$\varphi_v(-z) = (-1)^{v+1}\varphi_v(z). \tag{45}$$

Therefore, in each subband the average values of the z coordinate are the same, $\langle z \rangle_v = 0$. Consequently, in symmetric structures, the transition amplitude is independent of $B_\parallel$. The observed change in the maximum transition rate upon application of $B_\parallel$ is caused by the dependence of the transition amplitude $A^{rough}$ on the component of the magnetic field $B_\perp$ perpendicular to the layers of the structure. Applying a magnetic field parallel to the layers shifts the transition resonance to a different value of $B_\perp$. This also changes the transition amplitude $A^{rough}$ due to its dependence on $B_\perp$. Accordingly, this resonance shift also leads to a change in the value of scattering rate $1/\tau_{i\to f}$.

An analytical expression (24) is derived for the scattering amplitude, from which several general properties of scattering processes follow directly.

The transition amplitude is the product of two factors. The first factor $K_{\nu_f,\nu_i}$ is determined by the potential profile of the quantum well and is independent of the magnetic field. The second factor $Q_{n_i,n_f}(\gamma)$, in contrast, is dependent on the magnetic field.

The quantizing component of the magnetic field $B_\perp$ enters into formula (23) for the scattering amplitude through parameter $\gamma$ (31), which is the square of the ratio of the magnetic length $\ell_c = \sqrt{\hbar c/eB_\perp}$ to the interface correlation length $\lambda$.

The correlation length $\lambda$ can be viewed as the average size of a protruding island at the interface, while $\ell_c$ determines the localization area of the wave function in the plane of the layers (the standard deviation of the oscillator coordinate $\delta y = \ell_c \sqrt{n+1/2}$). Therefore, the fact that the scattering amplitude is determined by the ratio $\ell_c$ to $\lambda$ is quite natural. Furthermore, our expression shows that $B_\perp$ enters into the amplitude only through this ratio, and this dependence is universal, independent of the potential profile of the structure.

An important point is that the scattering amplitude depends not on the ratio itself, but on its square $\ell_c/\lambda$. This leads to the dependence of the scattering amplitude on the magnetic field varying significantly for different correlation lengths $\lambda$ (Fig. 5b). As a result, the nature of the dependence of the resonance shift on $B_\parallel$ is determined by the value of the average size of the roughness "island" $\lambda$ - depending on the value of $\lambda$, the resonance amplitude can either increase with $B_\parallel$, or decrease, or it can remain practically unchanged (Fig. 6). The obtained expressions allow us to explain this behavior.

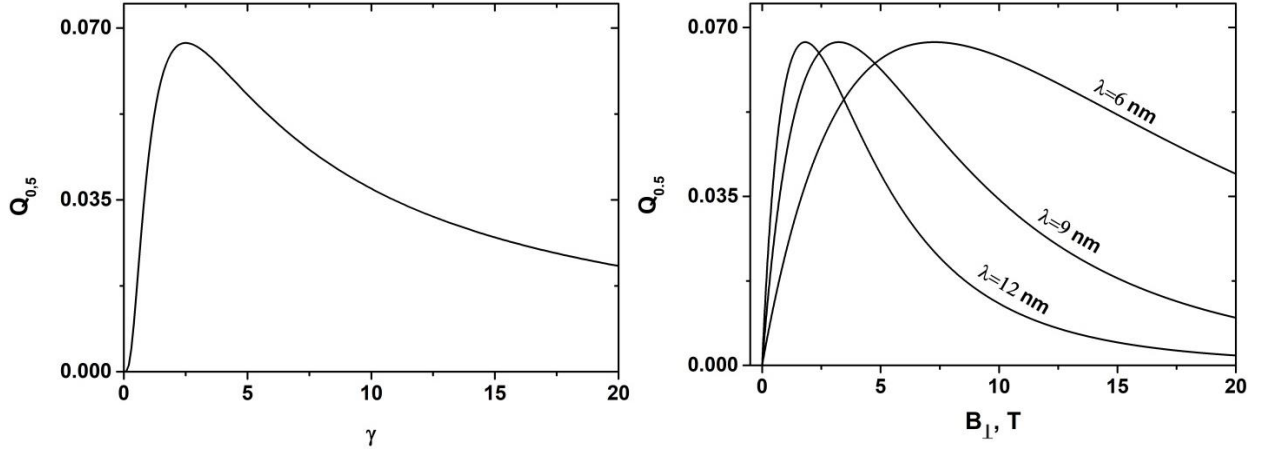

**Figure 5.** Dependence of the multiplier $Q_{0,5}$ on $\gamma$ (a) and on the magnetic field (b) for different values of the correlation length $\lambda$.

For small values of $\gamma$ (i.e., for $\ell_\perp \ll \lambda$) the multiplier $Q_{0,5}$ behaves as

$$Q_{n_i,n_f}(\gamma) = \binom{n_f}{n_i}(2\gamma)^{\Delta n} + o\left((2\gamma)^{\Delta n}\right). \tag{46}$$

At these values of $\gamma$, the size of the roughness "island" at the heterointerface is significantly larger than the size of the wave function localization area, and the electron weakly senses the roughness of the heterointerface. Consequently, scattering weakens.

In the opposite case of large values of $\gamma$ (i.e., when $\ell_\perp \gg \lambda$) $Q_{n_i,n_f}(\gamma)$ decreases with growth of $\gamma$ according to the law

$$Q_{n_i,n_f}(\gamma) = \frac{1}{2\gamma} - (n_i + n_f + 1)\frac{1}{(2\gamma)^2} + o\left(\frac{1}{(2\gamma)^2}\right). \tag{47}$$

This decrease in the scattering rate is caused by the wave function varying only slightly within the "roughness island."

Thus, as we move from one situation ($\ell_\perp \ll \lambda$) to another ($\ell_\perp \gg \lambda$), an increase in the scattering rate amplitude alternates with a decrease. Consequently, there is a maximum between these extreme cases (Fig. 7).

As follows from (46), with increasing $\Delta n$, the growth of the Q factor to the left of its maximum becomes less sharp, and, as a consequence, the position of the maximum shifts towards larger values of $\gamma$. At the same time, as comes from (47), the decrease of Q to the left of the maximum becomes slower with increasing $\Delta n$. For example, for the transition from the ground Landau level $(v_i, 0) \to (v_f, \Delta n)$, the maximum of $Q_{n_i, n_f}(\gamma)$ is achieved at

$$\gamma = \frac{\Delta n}{2}, \tag{48}$$

which corresponds to the magnitude of the magnetic field component $B_\perp$

$$B_{\perp, \Delta n}^{\max A} = \frac{2c\hbar}{e} \frac{1}{\lambda^2} \cdot \frac{1}{\Delta n}. \tag{49}$$

As can be seen, this value decreases with $\Delta n$ growth and significantly depends on the average size of the roughness "island" (Fig. 8).

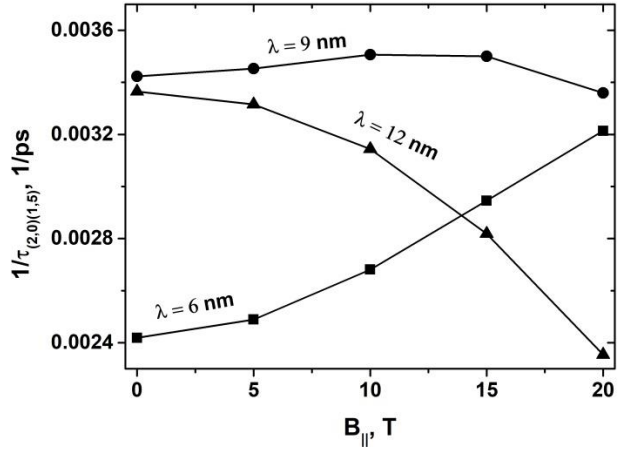

**Figure 6.** Dependence of the scattering rate $1/\tau_{i \to f}$ maximum on the magnetic field component $B_\parallel$ parallel to the layers for different values of the correlation length $\lambda$. The data are given for the transition $(2, 0) \to (1, 5)$ in a GaAs/Al$_{0.3}$Ga$_{0.7}$As quantum well of 25 nm width.

The ratio of the value of the magnetic field (33), at which the transition resonance occurs, to the value of the magnetic field (49), at which the maximum transition amplitude occurs,

$$\frac{B_{\perp, \Delta n}^{(0)}}{B_\perp^{\max A}} = \frac{1}{2} \frac{\Delta \varepsilon_{if}}{\hbar^2 / m_w \lambda^2} \tag{50}$$

does not depend on the change $\Delta n$ in the Landau level number during the transition, and is determined by the ratio of the intersubband distance and the energy of dimensional quantization of roughness $\hbar^2 / m_w \lambda^2$.

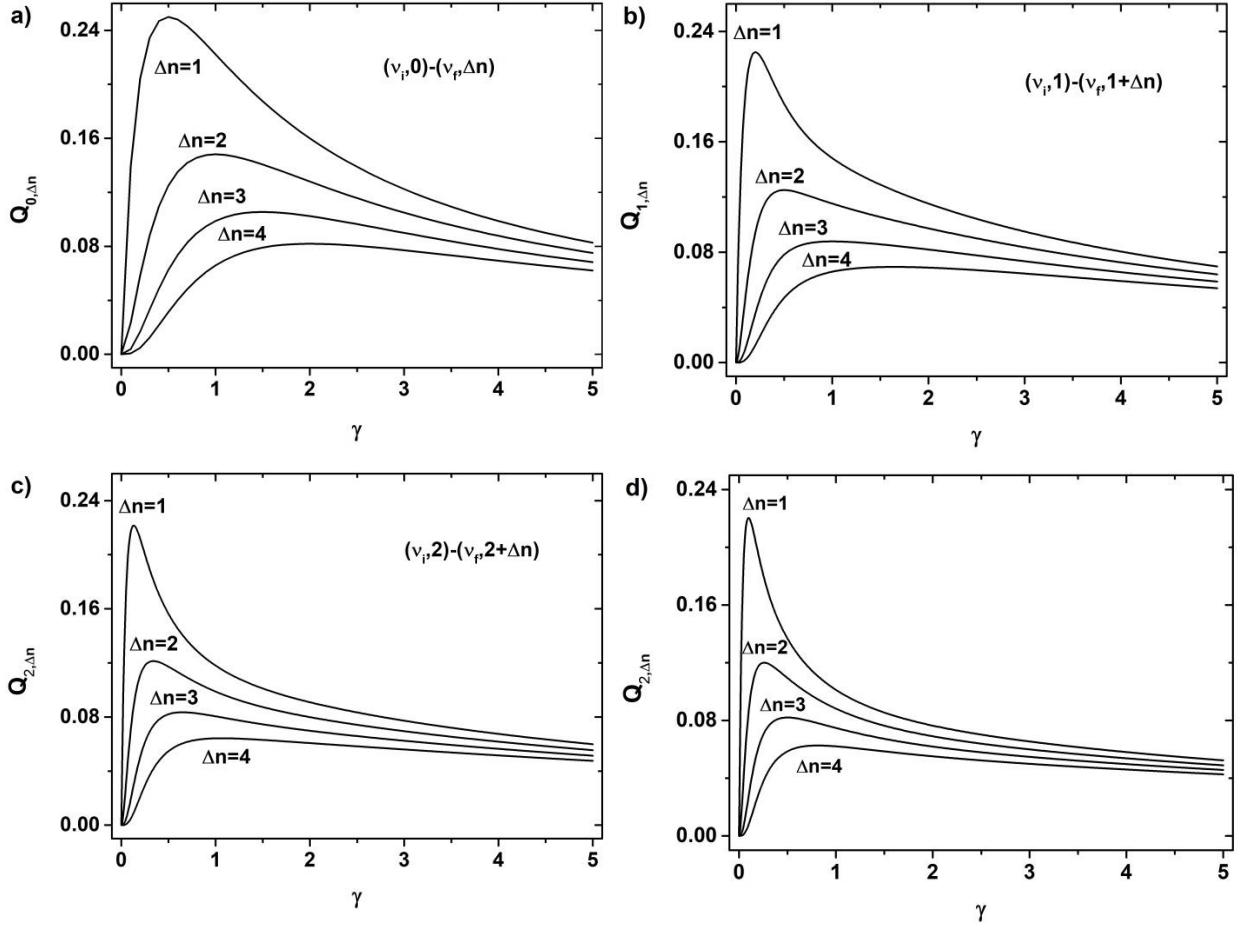

**Figure 7.** The multiplier $Q_{n_i, n_i+\Delta n}$ as a function of $\gamma$. a) $n_i = 0$; b) $n_i = 1$; c) $n_i = 2$; d) $n_i = 3$.

The position of the form factor maximum $B_{\perp,\Delta n}^{(0)}$ at $B_{\parallel} = 0$ coincides with the position of the maximum of the scattering amplitude at the correlation length value

$$\lambda_0 = 2\sqrt{\frac{\hbar^2}{2m_w \Delta \varepsilon_{if}}} \qquad (51)$$

At this point, the scattering rate reaches its maximum value.

If $\lambda > \lambda_0$ the resonance position at $B_{\parallel} = 0$ lies to the right of the maximum transition amplitude ($B_{\perp,\Delta n}^{(0)} > B_{\perp}^{\max A}$). In this case, the shift of the form factor maximum toward larger values of the quantizing component of the magnetic field $B_{\perp}$ with increasing $B_{\parallel}$ is accompanied by a decrease in the scattering amplitude $A^{rough}$ (Fig. 9c). Accordingly, the maximum scattering rate $1/\tau$ decreases (Fig. 6).

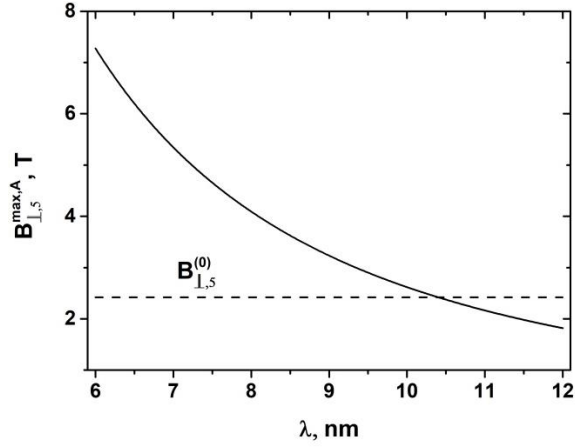

**Figure 8**. Dependence of the amplitude $A^{rough}$ of the interface roughness scattering on interface correlation length $\lambda$. The data are presented for the transition $(2,0) \rightarrow (1,5)$ in a 25 nm wide GaAs/Al$_{0.3}$Ga$_{0.7}$As quantum well. The dotted line shows the maximum form factor of this transition.

If $\lambda < \lambda_0$, the resonance position lies to the left of the transition amplitude maximum ($B^{(0)}_{\perp,\Delta n} < B^{\max A}_{\perp}$). In this case, the shift of the form factor maximum toward larger values of $B_\perp$ with increasing $B_\parallel$ is accompanied by an increase in the scattering amplitude $A^{rough}$ (Fig. 9a). Accordingly, the maximum scattering rate $1/\tau$ also increases (Fig. 6).

When the correlation length is close to $\lambda_0$, the resonance position passes through a maximum of $A^{rough}$ with increasing $B_\parallel$ (Fig. 9b). Accordingly, the maximum scattering rate initially increases with increasing $B_\parallel$, reaches a maximum value, and then decreases (Fig. 6).

Note that the maximum of $A^{rough}$ can always be reached (Fig. 10). The only question is at what value of $B_\parallel$. For relatively small values of $\lambda$, the maximum is reached at large values of $B_\parallel$ (Fig. 11). This value drops rapidly with increasing $\lambda$ and, at typical values of $\lambda$, reaches moderate magnetic fields $B_\parallel$ (several Tesla).

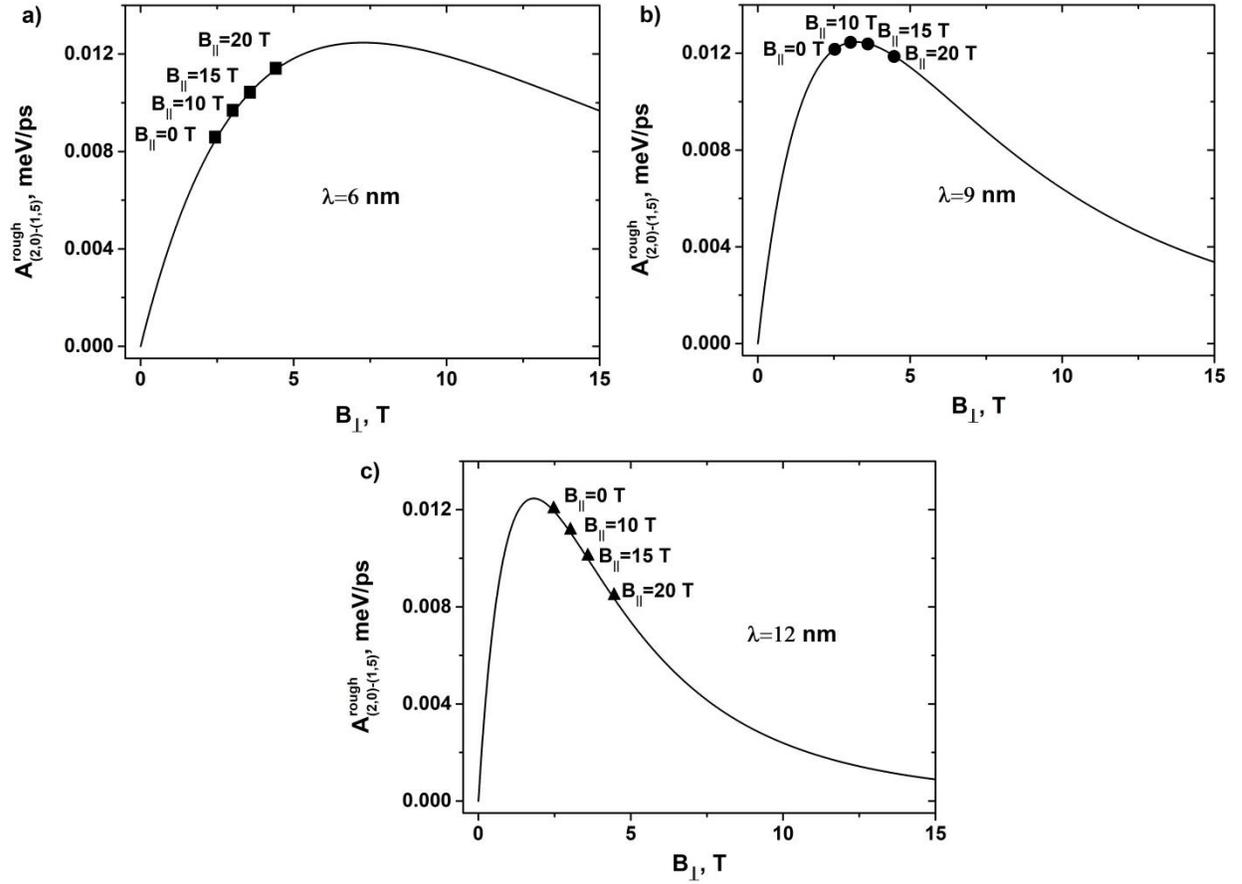

**Figure 9**. Scattering amplitude versus the quantizing component of the magnetic field $B_\perp$ in a symmetric quantum well for different correlation lengths $\lambda$. The data are presented for a transition $(2,0) \to (1,5)$ in a 25 nm wide GaAs/Al$_{0.3}$Ga$_{0.7}$As quantum well. The points show the positions of the form factor maxima for the corresponding value of the magnetic field component parallel to the quantum well layers.

Since the intersubband distance $\Delta\varepsilon_{if}$ decreases significantly with increasing quantum well width $a$, particularly for the lower levels of wide wells as $1/a^2$, similar behavior is observed with increasing quantum well width (Fig. 12). In relatively narrow quantum wells, the maximum of the scattering rate $1/\tau$ decreases with increasing $B_\parallel$. This decrease weakens with increasing quantum well width $a$, and starting from a certain value of $a$, it gives way to an increase, which intensifies with further increase of the quantum well width. The reason for this behavior is similar to that discussed above (Fig. 13).

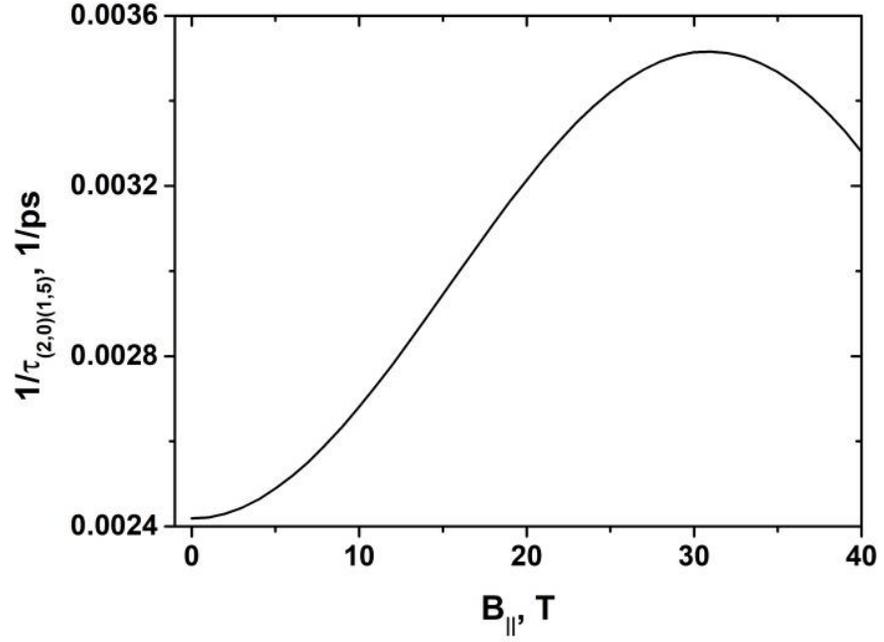

**Figure 10**. Dependence of the scattering rate maximum on the magnetic field component $B_\parallel$ parallel to the layers for the transition $(2,0) \to (1,5)$ in the $GaAs / Al_{0.3}Ga_{0.7}As$ quantum well of 25 nm width. Correlation length $\lambda = 6$ nm.

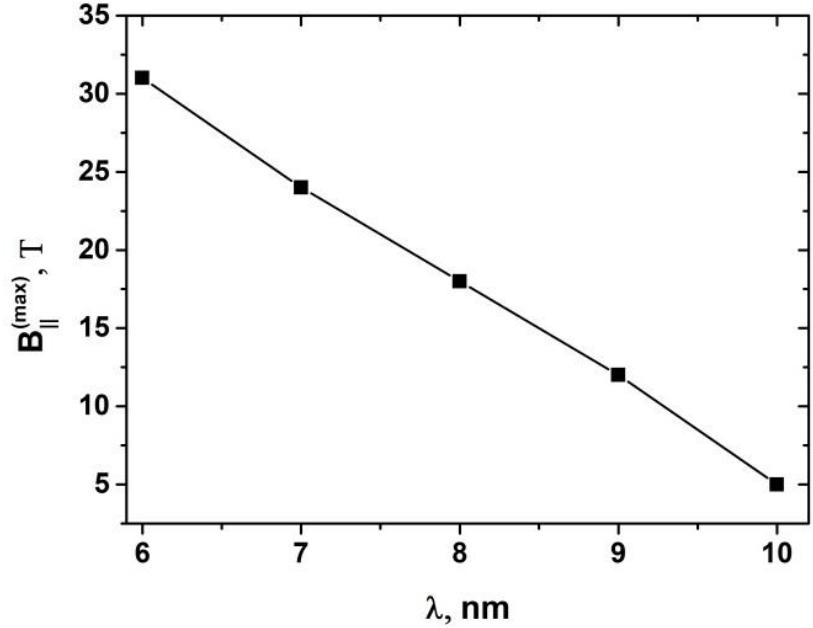

**Figure 11**. The value of the magnetic field component $B_\parallel^{(max)}$ at which the maximum transition rate is achieved as a function of the interface correlation length $\lambda$. The data are given for the transition $(2,0) \to (1,5)$ in the $GaAs / Al_{0.3}Ga_{0.7}As$ quantum well of 25 nm width.

To evaluate the law of such behavior, let us consider the deep levels in a rectangular quantum well. In this case,

$$\Delta\varepsilon_{if} = \frac{\hbar^2 \pi^2}{2m_w a^2}\left(v_i^2 - v_f^2\right), \quad (52)$$

and, correspondingly,

$$\frac{B_{\perp,\Delta n}^{(0)}}{B_\perp^{\max A}} = \frac{\pi^2\left(v_i^2 - v_f^2\right)}{4}\left(\frac{\lambda}{a}\right)^2 \quad (52)$$

Thus, the ratio of the magnetic field value $B_{\perp,\Delta n}^{(0)}$ at which the maximum form factor is achieved to the magnetic field value $B_\perp^{\max A}$ at which the maximum transition amplitude $A^{rough}$ is achieved is determined by the square of the ratio of the average roughness "island" size to the quantum well width. These two field values coincide if the quantum well has width

$$a_0 = \frac{\pi\sqrt{v_i^2 - v_f^2}}{2}\lambda \quad (53)$$

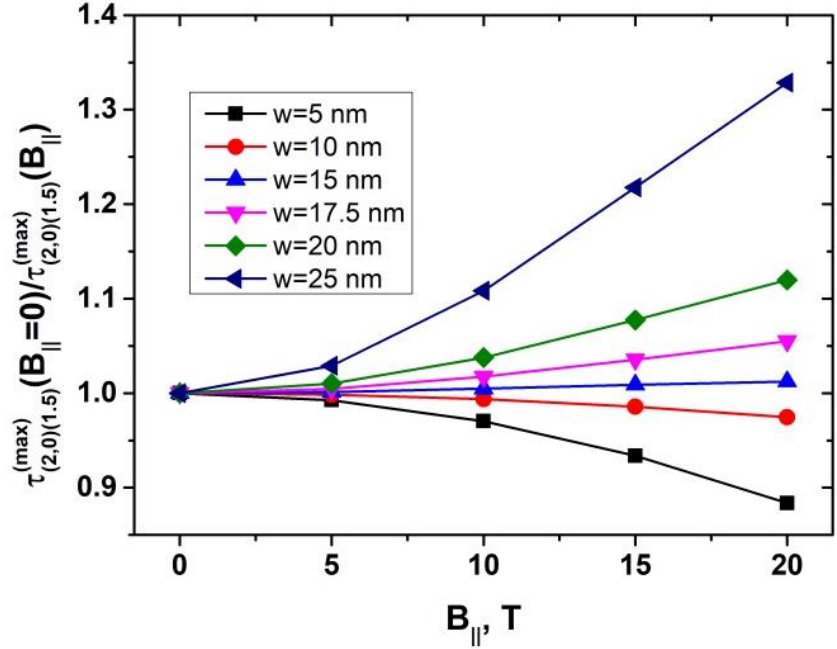

**Figure 12**. Dependence of the scattering rate $1/\tau$ at the maximum on $B_\parallel$ in quantum wells of different widths. The data are given for the transition $(2,0)\rightarrow(1,5)$ in the $GaAs/Al_{0.3}Ga_{0.7}As$ quantum well. Correlation length $\lambda = 6$ nm.

For narrower quantum wells ($a < a_0$) $B^{(0)}_{\perp,\Delta n} > B^{\max A}_{\perp}$, and, accordingly, the shift of the transition resonance toward higher magnetic fields leads to a decrease in the transition amplitude $A^{rough}$. Thus, in quantum wells with such a width, the maximum rate $1/\tau$ decreases monotonically with increasing the magnetic field component $B_\parallel$.

For quantum wells with width $a > a_0$ $B^{(0)}_{\perp,\Delta n} < B^{\max A}_{\perp}$. Therefore, initially the maximum rate $1/\tau$ increases with increasing $B_\parallel$. After reaching the maximum at

$$B_\parallel = \frac{2\sqrt{2}\pi\hbar c}{ea\lambda}\frac{v_i v_f}{\sqrt{v_i^2 - v_f^2}}\sqrt{1 - \frac{\pi^2(v_i^2 - v_f^2)}{4}\left(\frac{\lambda}{a}\right)^2}, \qquad (54)$$

the scattering rate $1/\tau$ decreases, first exceeding its value at $B_\parallel = 0$, and then becoming less than this value.

In structures with an asymmetric potential profile ($U(-z) \neq U(z)$), the wave functions $\varphi(z)$ of the quantum confinement levels are not divisible by parity. Therefore, the average coordinates in the states of different subbands $\langle z \rangle$ differ. Consequently, for $B_\parallel \neq 0$, the parameter $\xi$ for intersubband transitions is nonzero, and the transition amplitude is a function of not only $B_\perp$, but also of $B_\parallel$.

The dependence on $B_\parallel$ is manifested in the amplitude as a factor $G_{n_i,n_f}\left(\frac{\xi^2_{v_f,v_i}}{2(2\gamma+1)}\right)$, which is determined by expression (27). Note that $B_\parallel$ appears in the expression for the scattering rate amplitude only as a specific combination $x = \xi^2_{v_f,v_i}/2(2\gamma+1)$. In other words, the ratio of the transition amplitude in a tilted magnetic field to the transition amplitude at $B_\parallel=0$

$$G_{n_i,n_f}(x) = \frac{A^{rough}_{i \to f}(B_\parallel)}{A^{rough}_{i \to f}(B_\parallel = 0)} \qquad (55)$$

is a universal function of x, independent of the potential profile. This function is the product of a decaying exponential function $\exp(-x)$ and a polynomial $P_{n_i,n_f}(x)$. This polynomial is determined by the Landau level numbers of the initial and final states. For example, for the transition from the ground Landau level of the upper subband $(v_i, 0) \rightarrow (v_f, \Delta n)$, the polynomial has the form

$$P_{0,\Delta n}(x) = 1 + \sum_{j=1}^{\Delta n} \frac{1}{j!} \binom{\Delta n}{j} \frac{1}{(2\gamma)^j} x^j . \tag{56}$$

The coefficients of this polynomial depend on $\gamma$. This means that the behavior of $G(x)$ depends significantly on the value of $\gamma$—as $\gamma$ decreases, it transits from a monotonically decreasing function to a function with a maximum (Fig. 14).

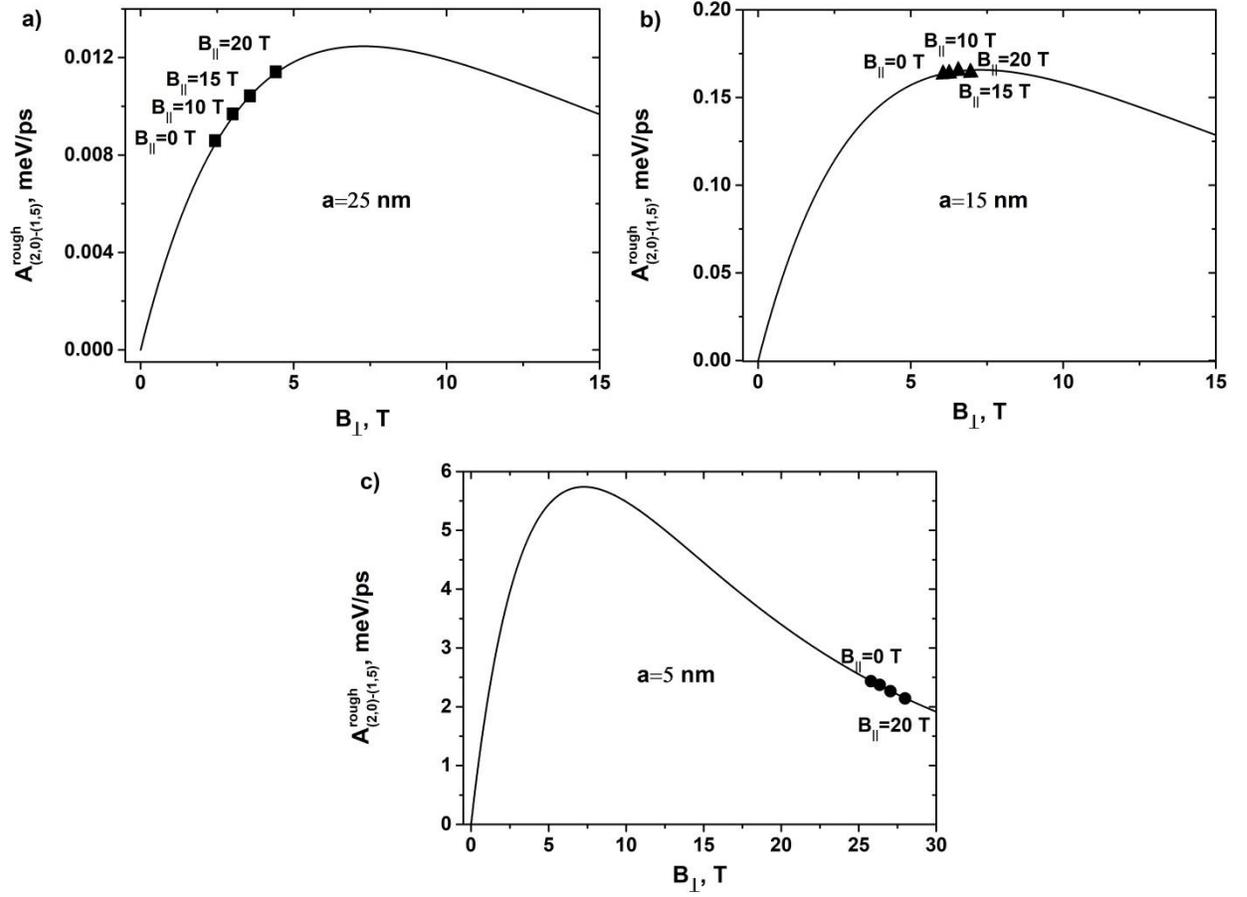

**Figure 13**. Scattering amplitude versus the quantizing component $B_\perp$ of the magnetic field in quantum wells of different widths. The data are presented for a transition $(2,0) \rightarrow (1,5)$ in a GaAs/Al$_{0.3}$Ga$_{0.7}$As quantum well. Correlation length = 6 nm. The points show the positions of the form factor maximum for the corresponding value of the magnetic field component $B_\parallel$ parallel to the quantum well layers.

For large values of the argument x, the function G(x) decreases exponentially:

$$G_{0,\Delta n}(x) \approx \frac{1}{\Delta n!}\frac{1}{(2\gamma)^{\Delta n}}\exp(-x)x^{\Delta n}, \qquad (57)$$

and the larger $\gamma$ the slower is this decrease.

For small x, the function G(x) behaves as

$$G_{0,\Delta n}(x) = 1 - \left(1 - \frac{\Delta n}{2\gamma}\right)x + o(x), \qquad (58)$$

For $1 - \frac{\Delta n}{2\gamma} > 0$, i.e., for $\gamma > \Delta n/2$, the function G(x) decreases for small x. This decrease continues for all x. This can be verified by calculating the derivative of the function G. This derivative has the form

$$G'_{0,\Delta n}(x) = -\exp(-x)\left[\begin{array}{l}\left(1 - \frac{\Delta n}{2\gamma}\right) + \sum_{j=1}^{\Delta n-1}\frac{1}{j!(2\gamma)^j}\binom{\Delta n}{j}\left\{1 - \left(\frac{\Delta n - j}{j+1}\right)\frac{1}{2\gamma}\right\}x^j + \\ + \frac{1}{\Delta n!(2\gamma)^{\Delta n}}x^{\Delta n}\end{array}\right]. \qquad (59)$$

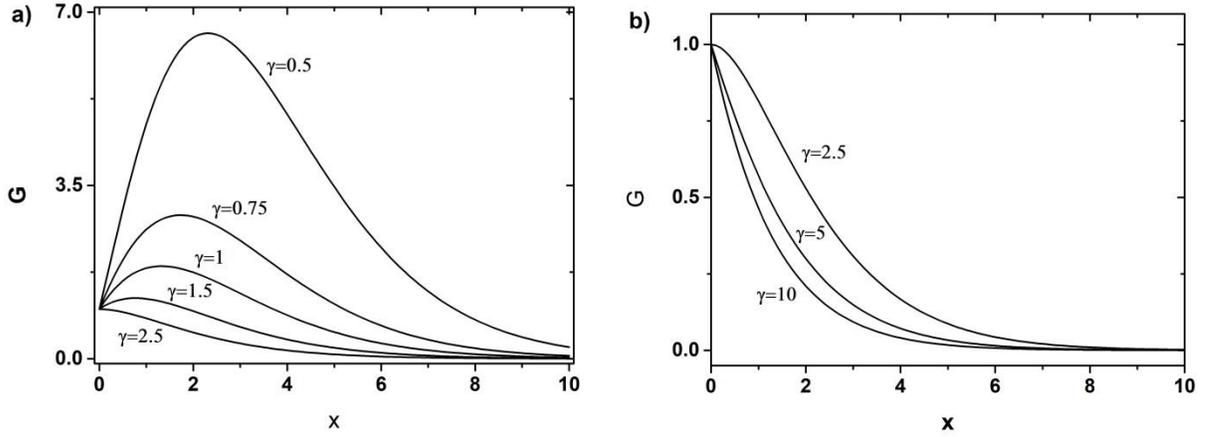

**Figure 14.** Ratio $G_{n_i,n_f}(x)$ of the scattering rate amplitude in a tilted magnetic field to the scattering amplitude at $B_\| = 0$, The data are given for the transition $(v_i,0) \to (v_f,5)$.

When the condition $1-\dfrac{\Delta n}{2\gamma}>0$ is satisfied, $1-\left(\dfrac{\Delta n-j}{j+1}\right)\dfrac{1}{2\gamma}>0$, and, accordingly, all terms in the square brackets of expression (58) are positive for all x. Consequently, for $\gamma>\Delta n/2$, the factor G(x) decreases monotonically for all x>0.

For $\gamma<\Delta n/2$, the function G(x) increases for small x. Since this function also decreases for large values of the argument, it reaches a maximum. Moreover, the slope of the tangent at x=0 decreases with increasing $\gamma$. Therefore, the larger $\gamma$, the larger the values of x at which this maximum is reached, and, accordingly, the smaller the magnitude of the factor G at the maximum. Accordingly, when $\gamma$ reaches the value of $\Delta n/2$, the factor G(x) becomes a monotonically decreasing function. In this case, the slope of the tangent at x=0 becomes negative and increases in absolute value. Therefore, the rate of decrease of the function G(x) increases with increasing $\gamma$.

## 5. Conclusion

Scattering processes by the interface roughness in a quantum well in a quantizing magnetic field are considered. Expressions for the scattering rate in a magnetic field tilted relative to the plane of the quantum well layers are derived. These expressions are analyzed for transitions of various types, and trends in the behavior of the scattering rate with changes in the magnetic field strength and orientation are established, as well as the influence of the quantum well potential profile and roughness parameters on the scattering rate.

It is shown that the magnetic field component parallel to the layers has virtually no effect on intrasubband scattering processes.

Two aspects can be distinguished in the influence of the magnetic field component $B_{\|}$ on intersubband transitions.

First, the component $B_{\|}$ leads to an increase in the distance between subbands. This increase in the intersubband distance leads to a shift in the resonance position toward higher values of the quantizing magnetic field component $B_{\perp}$ directed perpendicularly to the quantum well layers. This shift depends significantly on the quantum well width $a$ – it increases with increasing well width, approximately proportionally to $a^2$.

The second aspect of the effect of $B_\parallel$ on scattering processes is the influence of the magnetic field on the electron wave function and, consequently, on the transition amplitude $A^{rough}$.

The magnitude of the effect depends significantly on the symmetry of the quantum well potential profile U(z).

In symmetric quantum wells ($U(-z) = U(z)$), the scattering amplitude is independent of $B_\parallel$. Therefore, in such structures, the main effect is a shift of the resonance toward larger values of the quantizing component of the magnetic field $B_\perp$, which is accompanied by a change in the maximum transition rate due to the dependence of the scattering amplitude $A^{rough}$ on $B_\perp$. An analytical expression for this dependence is obtained. It is shown that the transition amplitude is the product of two factors. The first factor is determined by the quantum well potential profile and is independent of the magnetic field. The second factor, conversely, is determined only by the magnetic field and is independent of the quantum well potential profile. It is shown that the quantizing component of the magnetic field is included in the transition amplitude only as a combination of $\gamma = (\ell_c / \lambda)^2$. The dependence $A^{rough}$ on $\gamma$ is nonmonotonic: an increase of $A^{rough}$ with increasing $\gamma$, followed by a decrease $A^{rough}$ upon reaching a maximum. The maximum of this dependence coincides with the value of $B_\perp$ at the transition resonance only for a certain combination of the quantum well width and the heterointerface correlation length. The nature of the dependence of the scattering rate amplitude on $B_\parallel$ is determined by the ratio of the quantum well width and the correlation length.

In asymmetric quantum wells, the scattering amplitude depends on both $B_\perp$, and $B_\parallel$.

It is shown that the ratio G of the transition amplitude in a tilted magnetic field to the transition amplitude in a perpendicular field is a universal function of the parameter $x = \xi^2_{v_f, v_i} / (2\gamma + 1)$. The behavior of G(x) depends on $\gamma$. For small $\gamma$, the dependence G(x) is nonmonotonic: it initially increases, reaches a maximum, and then decreases. At higher values of $\gamma$, the shape of the function G(x) changes, it gradually transforms from a resonant peak to a monotonically decreasing function.